\documentclass[traditabstract]{aa}
\usepackage{natbib}
\usepackage{epsfig}
\usepackage{txfonts}

\usepackage{graphicx}
\usepackage{txfonts}

\usepackage{epsf}
\usepackage{rotating}
\usepackage{graphics}
\usepackage[final]{pdfpages}

\def\ms{\,m\,s$^{-1}$}         
\def\kms{\,km\,s$^{-1}$}         
\def\ms{\hbox{\,m\,s$^{-1}$}}         
\def\m2s2{\hbox{\,m$^{2}$\,s$^{-2}$}} 
\def\kms{\hbox{\,km\,s$^{-1}$}}       
\def\vsini{\hbox{$v$\,sin\,$i_s$}}      
\def\Msun{\hbox{M$_{\odot}$}}             

\def\Mjup{\hbox{$\mathrm{M}_{\rm Jup}$}}
\def\Rjup{\hbox{$\mathrm{R}_{\rm Jup}$}}

\def \mp{$M_{\rm p}$}
\def \rp{$R_{\rm p}$}
\def\8{HAT-P-8}
\def\9{HAT-P-9}
\def \2{HAT-P-23}
\def \6{HAT-P-16}

\def \1s{$1\,\sigma$}

\def \t0{T$_0$}
\def \mp{$M_{\rm p}$}

\begin{document}

\title{Spin-orbit inclinations of the exoplanetary systems \8, \9\, \6, and \2 \thanks{Based on observations collected with the SOPHIE spectrograph on the 1.93-m telescope at Observatoire de Haute-Provence (CNRS), France, by the SOPHIE Consortium (program 10A.PNP.CONS).}}

\author{
Moutou,~C. \inst{1}
\and D\'iaz, R. F. \inst{2,3}
\and Udry,~S. \inst{4}
\and H\'ebrard,~G. \inst{2,3}
\and Bouchy,~F. \inst{2,3}
\and Santerne, A. \inst{1,3}
\and Ehrenreich,~D. \inst{5}
\and Arnold, L.  \inst{3}
\and Boisse,~I. \inst{2,6}
\and Bonfils,~X. \inst{5}
\and Delfosse,~X. \inst{5}
\and Eggenberger,~A. \inst{5}  
\and Forveille,~T. \inst{5} 
\and Lagrange,~A.-M. \inst{5}
\and Lovis,~C. \inst{4}
\and Martinez, P. \inst{7}
\and Pepe,~F. \inst{4}
\and Perrier,~C. \inst{5} 
\and Queloz,~D. \inst{4}
\and Santos,~N.C. \inst{6,10}
\and S\'egransan,~D. \inst{4}
\and Toublanc, D. \inst{7,8}
\and Troncin,~J.P. \inst{3}
\and Vanhuysse,~M. \inst{9}
\and Vidal-Madjar,~A. \inst{2} 
}
\offprints{C. Moutou}

\institute{
Laboratoire d'Astrophysique de Marseille, UMR 6110, OAMP, CNRS\&Univ. de Provence, 38 rue Fr\'ed\'eric Joliot-Curie, 13388 Marseille cedex 13, France,
\email{Claire.Moutou@oamp.fr}
\and Institut d'Astrophysique de Paris, UMR 7095 CNRS, Universit\'e Pierre \& Marie Curie, 98bis boulevard Arago, 75014 Paris, France
\and Observatoire de Haute-Provence, CNRS \& OAMP, 04870 Saint-Michel l'Observatoire, France
\and Observatoire de Gen\`eve, Universit\'e de Gen\`eve, 51 Chemin des Maillettes, 1290 Sauverny, Switzerland
\and UJF-Grenoble 1 / CNRS-INSU, Institut de Plan\'etologie et d'Astrophysique de Grenoble (IPAG) UMR 5274, Grenoble, F-38041, France 
\and Centro de Astrof\'isica, Universidade do Porto, Rua das Estrelas, 4150-762 Porto, Portugal
\and Association Adagio, L'Observatoire, 31540 Belesta Lauragais, France
\and Universit\'e Toulouse III, UMR5187, 118 route de Narbonne 31062 Toulouse, France
\and Oversky, 47 All\'ee des Palanques, 33127 Saint Jean d'Illac, France
\and Departamento de F\'isica e Astronomia, Faculdade de Ci\^encias, Universidade do Porto, Portugal
}
 
  \abstract{  
We report the measurement of the spin-orbit angle of the extra-solar planets  \8 b, \9 b, \6 b and \2 b,  thanks to spectroscopic observations performed 
at the Observatoire de Haute-Provence with the SOPHIE spectrograph on the 1.93-m 
telescope.  Radial velocity measurements of the Rossiter-McLaughlin effect show the 
detection of an apparent prograde, aligned orbit for all systems.
The projected spin-orbit angles are found to be $\lambda=$-17$^\circ$$^{+9.2}_{-11.5}$, -16$^{\circ}$ $\pm$ 8$^{\circ}$, 
-10$^{\circ}$ $\pm$ 16$^{\circ}$, +15$^{\circ}$ $\pm$ 22$^{\circ}$ for \8, \9, \6 and \2, respectively, with corresponding projected rotational velocities of 14.5 $\pm$ 0.8, 12.5 $\pm$ 1.8, 3.9 $\pm$ 0.8, and 7.8 $\pm$ 1.6  \kms. These new results increase to 37 the number of accurately measured spin-orbit angles in transiting extrasolar systems. We conclude by drawing a tentative picture of the global behaviour of orbital alignement, involving the complexity and diversity of possible mechanisms. 
}
\titlerunning{Spin-orbit inclinations of four exoplanetary systems}
\authorrunning{C. Moutou et al.}

\maketitle

\section{Introduction}
The number of discovered extrasolar planetary systems is constantly increasing, at a rate larger than 50 per year since 2007 \footnote{www.encyclopedia.edu}. The sub-sample of confirmed transiting systems ($\sim$ 100) is explored with greater care, since the measurement of their real mass and radius allows studying their internal structure. For a sub-group of transiting systems, several teams have now collected the observation of spectroscopic transits with high-precision radial-velocity instruments, giving access to the relative angle between the stellar spin plane and the planetary orbital plane through the modeling of the Rossiter-McLaughlin effect \citep{rossiter}.  Unexpectedly, one third of the short-period planets for which those measurements are available have significantly misaligned orbits with respect to their star's rotational axis, even polar or retrograde  (e.g. \citet{triaud10,hebrard10,winn10}). Such a distribution of stellar obliquities requires a scenario of formation and orbital evolution of extrasolar giant planets more diverse than pure one-planet migration inside a proto-planetary disk.

In this paper, we report the Rossiter-McLaughlin measurement of four additional transiting planets, \8 b, \9 b, \6 b, \2 b, obtained at Observatoire de Haute Provence with the SOPHIE high-precision radial-velocity (RV) spectrograph. The four systems discussed here are transiting extrasolar planets found by  the HATNet photometric survey \citep{bakos02,bakos04}. 

\8 b is an inflated hot Jupiter, with a mass \mp = 1.52 \Mjup, a radius \rp = 1.50 \Rjup\, and a circular orbit of period 3.076 days \citep{l08}. The host star \8 is a solar-metallicity F main-sequence star with a significant projected velocity of about 12 km/s, and magnitude $V=$10.17. 

\9 is an inflated, moderate-mass Jupiter-like planet, with \mp=0.78 \Mjup\ and \rp=1.4 \Rjup. It orbits a late F star in a 3.92289-day circular orbit \citep{shporer09}. The star is relatively faint, with $V=12.3$. Ephemeris have been recently revisited \citep{dittmann10}; we used the updated value for the observation scheduling and transit modeling.

\6 b is  a 4 \Mjup\  planet transiting a $V=$10.8 magnitude star; its orbital period is 2.77596 days, its radius is 1.29 \Rjup\ and it has a slightly eccentric orbit despite the short period (e=0.036 with a 10$\sigma$ significance). The parent star has a mass of 1.2 \Msun\, and an effective temperature of  6158K \citep{buchhave10}.

\2 b is an inflated and massive giant planet orbiting a G0 dwarf star with $V=11.94$ , in 1.212884d.  Its mass and radius are 2.09 \Mjup\  and 1.37 \Rjup; although the orbital period is extremely short, the orbit of \2 b is slightly but significantly eccentric, with $e=$0.106. Its detection and analysis have been reported by \citet{bakos2011b}.

Section \ref{spec} describes the spectroscopic observations, section 3 presents additional photometry that was performed and analysed to refine the ephemeris of \8 b, section \ref{mod} discusses the modeling of the Rossiter-McLaughlin effect. In section \ref{disc} we put these discoveries into context and discuss the global picture of orbital alignment.

\section{Spectroscopic observations}
\label{spec}

The spectroscopic observations have been obtained at Observatoire de Haute Provence with the 1.93m telescope, in the framework of the large program (reference 10A.PNP.CONS) led by the SOPHIE Consortium, presented in detail by \citet{bouchy2009}. The SOPHIE instrument \citep{perruchot} is a fiber-fed environmentally stabilized echelle spectrograph covering the visible range from 387 to 694 nm. The spectral resolving power is 70,000 in the High-Resolution mode, and 40,000 in the High-Efficiency mode. For high-cadence observations required in a Rossiter sequence, the limit between both modes is around $V=9.5-10$: the gain in resolution does not compensate the increase of readout noise for fainter stars.

The observing template "objAB" is used, in order to simultaneously monitor the sky background contamination. 
The  correction is required when the sky brightness is large and the Earth velocity is not sufficiently separated from the stellar velocity.

The radial velocities are obtained by cross-correlating the 2D extracted spectra with a stellar numerical mask, following the method developed by \citet{baranne}. 

\subsection{\8}
The transit of \8\, b has been observed during the night of 19 July 2010. We observed the $V=$ 10.17 mag \8 star in the High-Efficiency mode, and slow readout mode to reduce the detector noise. Three measurements of this target were collected in 2009, and three in 2010 outside the transit night.  These few measurements permitted to combine with a properly defined offset the orbital radial-velocity variations to the set of higher-accuracy Keck/HIRES data described in \citet{l08}. During the transit night (around JD=55397.5), we secured 35 measurements, between 21:40 to 02:40 UT. Their signal-to-noise ratios range from 37 to 42, with exposure times from 220 to 640 seconds. Observing at constant signal-to-noise ratio with SOPHIE allows reducing the radial-velocity systematics, that are due to the illumination-sensitive Charge Transmission Efficiency \citep{bouchy09}. Although this systematic effect can be efficiently corrected after the proper calibration effect has been quantified, by observing at constant signal-to-noise ratio we reduce the effect. The individual error bars range from 16 to 21 m/s. The six bluest orders were removed for the velocity calculation, since they only add noise, and the G2 mask was used for the cross-correlation.
The 1-$\sigma$ error bar of our measurements within transit is 18 m/s. It is similar for the measurements collected outside transit.  

\begin{figure}
\begin{center}
\hspace*{-0.8cm}
\epsfig{file=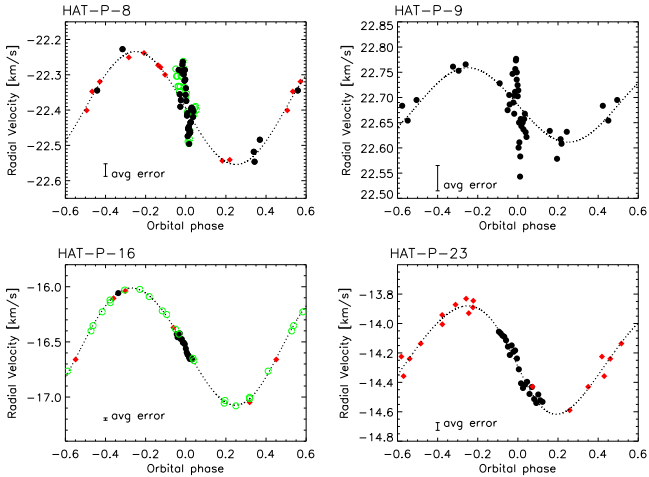,width=0.55\textwidth,angle=180}
\caption{SOPHIE (black plain circles), Keck/HIRES (red diamonds), FIES (open green circles) radial-velocity data of \8, \9, \6, and \2. The superimposed orbital solution is our best-fit to the data (not including the spectroscopic transits in this plot).}
\label{orbit}
\end{center}
\end{figure}

The RV data of \8 presented in this paper are shown on Figure \ref{orbit}, together with the previous Keck/HIRES data of \citet{l08}. They are phased to the ephemeris refined by our photometric observations (see section 3). Modeling the full out-of-transit radial-velocity sequence with both  Keck and  SOPHIE data with a Keplerian solution, we find a semi-amplitude of the variation 1.5-$\sigma$ larger than in \citet{l08}, and values for the planetary mass and the semi-major axis that are varied accordingly. We update the orbital parameters and planetary mass in Table \ref{param}. The eccentricity being compatible with 0 at less than 2-$\sigma$ level, we assume here a circular orbit, as in \citet{l08}. The bisector span is constant to the 2-$\sigma$ level during the sequence.

The velocity anomaly during the transit has an amplitude only 30\% smaller than the orbital amplitude; this was expected from the significant projected velocity of the star and the large planetary radius  (11.5 km/s and 1.5 \Rjup\ respectively, in \citet{l08}). This \vsini\ can also be estimated from the cross-correlation function, as described in \citet{boisse10}. We find \vsini = 12.9 $\pm$ 1 km/s, larger but in agreement with the value reported in \citet{l08}.

\begin{table}
\centering
  \caption{Updated orbit and planetary parameters for \8\,b. A circular orbit is assumed.}
  \label{param}
\begin{tabular}{lcc}
\hline
\hline
Parameters & Values and 1-$\sigma$ error bars & Unit \\
\hline
$V_r$  (SOPHIE)               & -22.383 $\pm $   0.01                   &   \kms        \\
$P$                                     & 3.0763373 $\pm$    0.0000031                    &   days        \\
$K$                                    & 158.2 $\pm$  3.2                            &   \ms \\
$M_\mathrm{p} $    &   1.34$\pm$0.05           &   M$_\mathrm{Jup}$ \\
$a$                                    &   0.0449 $\pm$ 0.0007       &  AU \\
$T_t$ (primary transit)    & 2454437.67505 $\pm$ 0.00042&   BJD         \\
\hline
\end{tabular}
\end{table}

\subsection{\9}
Two partial transits of \9 have been recorded, on 28 December 2010 and 1 January 2011 in the High-Efficiency mode and slow readout. Ten (seventeen) measurements were secured during the first (resp. second) night. Exposure times ranging from 830 to 1200 seconds resulted in spectra with signal-to-noise ratio ranging from 14 to 31. An error of 12 m/s was quadratically added to the photon noise uncertainty, to account for instrumental systematics at low signal-to-noise ratio and get a final $\chi^2$ value close to 1 for the orbital fitting.
We added the fifteen measurements obtained with SOPHIE in 2007-2008, at the planet's discovery by \citet{shporer09}. All spectra were reduced in the same way, using the F0 correlation mask and spectrograph orders 3 to 34. The values are reported in Table \ref{rv9}.  We finally adjusted the planetary orbit using the last ephemeris derived by additional photometry by \citet{dittmann10}. All RV data are shown on Figure \ref{orbit}.

The transit-center epochs, using this ephemeris, are expected at Julian dates 5559.44531 and 245563.37109. Due to variations of the weather conditions, we had to interrupt the first sequence 30 minutes before the end of the transit, and we did not get data after transit on the 28 December 2010 (observations lasted from 20:17 to 23:54 UT). The second sequence nicely completes the first one, with a full transit observed and seven measurements after transit (from 19:20 to 00:59 UT). No data were obtained before transit however, because of the twilight. The radial-velocity time series of both transits were assembled by letting a free offset between both sequences, to allow for an observed shift of 37 m/s. The reason for this may be related to the atmospheric chromatic dispersor setup. This, however, does not cause additional jitter during the short timescale of the transit itself, thus all data are kept in the fit. The bisector span is constant at the 1.6-$\sigma$ level during the sequence.
The average width of the cross-correlation function is 16.6 km/s, which corresponds to a projected rotational velocity of 12.2 km/s, with a precision of about 1 km/s. This value is compatible with the one found by \citet{shporer09} from the analysis of stellar parameters (11.9 $\pm$ 1 km/s) .

\subsection{\6}
The transit of \6\, b has been observed during the night of 15 August 2010 with SOPHIE in the High-Resolution mode and fast readout mode. In order to maintain the signal-to-noise ratio constant ($\sim 27 \pm 1$) over the 5-hour sequence, an exposure time of 848 to 1400 s has been used. The cross-correlation is calculated with a G2 mask which lowers the individual error bars, excluding the 14 bluest orders to diminish the systematic effects due to an incorrect positioning of the atmospheric chromatic dispersor during this observation. A RV precision of 10 to 15 m/s has been obtained. The stellar spectrum was not affected by any significant background light. One additional measurement was obtained the night before the transit in order to assess the RV uncertainty. Eighteen measurements were secured during the transit night. The sequence lasted from 22:19 to 03:27 UT. The transit was expected to occur between 23:47 and 02:51 UT, following ephemeris in \citet{buchhave10}.The amplitude of the RV anomaly is extremely small, and the pattern appears regular as for a prograde aligned orbit, with a redshift followed by a blueshift. Figure \ref{orbit} shows the SOPHIE data overplotted with literature data obtained with FIES and HIRES \citep{buchhave10}. The bisector span is constant within 1.4-$\sigma$ during the sequence.

A projected stellar rotational velocity of 3.0 $\pm$ 1 km/s is found from fitting the cross-correlation function, in agreement with the value reported in \citet{buchhave10} (3.5 $\pm$ 0.5 km/s)).

\subsection{\2}

The transit of \2\, b has been observed during the night of 25 August 2010, a few days after its announcement by \citet{bakos2011}. In this sequence, SOPHIE was used in the High-Efficiency mode and fast readout mode. In order to maintain the signal-to-noise ratio constant ($\sim 24 \pm 1$) over the 6.3-hour sequence, an exposure time of 787 to 1200 s has been used. A RV precision of 24 to 36 m/s has been obtained, after cross correlating with the G2 mask, including all orders. Due to the full Moon, there is a significant background contamination entering the spectrograph fibers. The stellar spectrum is a posteriori corrected for this effect, using the neighbour fiber monitoring the sky flux simultaneously to each exposure. The value of the correction varies from 100 to 500 m/s during the sequence, increasing towards the end of the sequence when the moon light became more prominent. The velocity difference between the barycentric Earth RV and the target's RV is about 5 km/s during the transit night. This contamination by the background light induces a residual jitter in the last data points of the series, but does not prevent the detection of the RM anomaly, as seen on Figure \ref{orbit}. 
The sequence on 25 August lasted from 20:01 to 02:20 UT, with a transit expected from 21:41 to 23:51 UT following ephemeris of \citet{bakos2011b}.
Twenty-three measurements were obtained during the transit night, plus one additional measurement two nights before the transit. The amplitude of the RM anomaly is of the order of 100m/s and the orbit is prograde. Keck/HIRES radial velocities were added to our data set in order to simultaneously fit the orbit and the transit (Figure \ref{orbit} and section \ref{mod}). The bisector span is constant within 1$\sigma$ during the sequence.

We measure a projected stellar rotational velocity of 10 $\pm$ 1 km/s, larger by 2-$\sigma$ than the value of 8.1 $\pm$ 0.5 km/s reported in \citet{bakos2011b}.

\onltab{1}{
\begin{table}[b]
\centering
  \caption{Radial velocity measurements obtained with SOPHIE of \8.}
  \label{rv8}
\begin{tabular}{lcc}
\hline
JD-2,400,000.  &  Radial Vel. & Uncertainty \\
         & [km s$^{-1}$]   &  [m s$^{-1}$] \\
\hline
55072.45560  & -22.5463  &   18.4 \\
55080.52188  & -22.2864  &   18.5 \\
55088.49962  & -22.3445  &   18.9 \\
55395.55227  & -22.4840  & 20.8 \\
55396.52298  & -22.2275  &   17.8 \\
55397.39997  & -22.3553  &   18.5 \\
55397.40930  & -22.3904  &   18.0 \\
55397.41822  & -22.3718  &   18.6 \\
55397.43378  & -22.2944  &   17.6 \\
55397.44022  & -22.2714  &   17.5 \\
55397.44685  & -22.2634  &   17.5 \\
55397.45327  & -22.2887  &   18.3 \\
55397.46034  & -22.2989  &   17.8 \\
55397.46719  & -22.2869  &   17.2 \\
55397.47357  & -22.3169  &   17.3 \\
55397.47896  & -22.2864  &   17.1 \\
55397.48420  & -22.3135  &   17.1 \\
55397.48938  & -22.3570  &   17.5 \\
55397.49458  & -22.3382  &   17.4 \\
55397.50028  & -22.3741  &   17.3 \\
55397.51130  & -22.4156  &   17.5 \\
55397.51675  & -22.4090  &   17.5 \\
55397.52222  & -22.4740  &   17.7 \\
55397.52780  & -22.4527  &   17.8 \\
55397.53327  & -22.4539  &   17.6 \\
55397.53856  & -22.4475  &   17.8 \\
55397.54369  & -22.4959  &   17.6 \\
55397.54868  & -22.4662  &   17.7 \\
55397.55407  & -22.4575  &   17.5 \\
55397.55914  & -22.4639  &   17.6 \\
55397.56427  & -22.4359  &   18.0 \\
55397.56930  & -22.4345  &   18.1 \\
55397.57425  & -22.3988  &   18.1 \\
55397.57918  & -22.3946  &   18.4 \\
55397.58403  & -22.3932  &   18.5 \\
55397.58884  & -22.3961  &   18.3 \\
55397.59440  & -22.3973  &   18.6 \\
55397.59935  & -22.3989  &   18.5 \\
55397.60439  & -22.4194  &   18.4 \\
55397.60947  & -22.4037  &   18.4 \\
55398.53561  & -22.5182  &   15.5 \\
\hline
\end{tabular}
\end{table}
}

\onltab{2}{
\begin{table}[b]
\centering
  \caption{Radial velocity measurements obtained with SOPHIE of \9.}
  \label{rv9}
\begin{tabular}{lcc}
\hline
JD-2,400,000.  &  Radial Vel. & Uncertainty \\
         & [km s$^{-1}$]   &  [m s$^{-1}$] \\
\hline
55559.34192 &22.6812& 36.9\\
55559.35866 &22.5790 &43.2\\
55559.37506 &22.5852 &33.5\\
55559.39135 &22.6719 &35.1\\
55559.40763 &22.6588 &34.0\\
55559.42391 &22.6997 &28.8\\
55559.44366 &22.6173 &34.0\\
55559.46000 &22.5507 &37.2\\
55559.47627 &22.5503 &35.2\\
55559.49255 &22.5119 &56.9\\
55563.30211 &22.6347 &32.3\\
55563.31818 &22.6929 &33.5\\
55563.33464 &22.7358 &27.9\\
55563.34868 &22.7057 &27.4\\
55563.36333 &22.6758 &28.9\\
55563.37770 &22.6792 &28.5\\
55563.39233 &22.6012 &29.1\\
55563.41194 &22.5295 &28.1\\
55563.42434 &22.5684 &28.4\\
55563.43747 &22.6301 &31.1\\
55563.45125 &22.6167 &26.0\\
55563.46384 &22.5940 &26.2\\
55563.47781 &22.6040 &25.1\\
55563.49242 &22.6422 &24.5\\
55563.50598 &22.6353 &24.3\\
55563.52041 &22.5892 &23.8\\
55563.53780 &22.5987 &23.7\\
\hline
\end{tabular}
\end{table}
}

\onltab{3}{
\begin{table}[b]
\centering
  \caption{Radial velocity measurements obtained with SOPHIE of \6.}
  \label{rv16}
\begin{tabular}{lcc}
\hline
JD-2,400,000.  &  Radial Vel. & Uncertainty \\
         & [km s$^{-1}$]   &  [m s$^{-1}$] \\
\hline
55424.43002&	-16.4335&	10.7Ê\\
55424.44167&	-16.4491&	10.4Ê\\
55424.45587&	-16.4358&	12.5Ê\\
55424.47137&	-16.4391&	15.4Ê\\
55424.48567&	-16.5030&	15.4Ê\\
55424.50068&	-16.4913&	14.5Ê\\
55424.51498&	-16.5081&	11.6Ê\\
55424.52851&	-16.5096&	10.8Ê\\
55424.54198&	-16.5225&	11.2Ê\\
55424.55597&	-16.5680&	11.3Ê\\
55424.56858&	-16.5967&	11.0Ê\\
55424.57975&	-16.6180&	10.9Ê\\
55424.59027&	-16.6271&	10.9Ê\\
55424.60097&	-16.6502&	10.9Ê\\
55424.61203&	-16.6542&	11.1Ê\\
55424.62317&	-16.6594&	10.8Ê\\
55424.63377&	-16.6644&	10.9Ê\\
55424.64422&	-16.6681&	10.8Ê\\
\hline
\end{tabular}
\end{table}
}

\onltab{4}{
\begin{table}[b]
\centering
  \caption{Radial velocity measurements obtained with SOPHIE of \2.}
  \label{rv23}
\begin{tabular}{lcc}
\hline
JD-2,400,000.  &  Radial Vel. & Uncertainty \\
         & [km s$^{-1}$]   &  [m s$^{-1}$] \\
\hline
55434.33411	&-14.0555	&24.5 \\
55434.34457	&-14.0666	&25.6 \\
55434.35558	&-14.0820	&24.5 \\
55434.36645	&-14.0898	&24.9 \\
55434.37731	&-14.1137	&25.2 \\
55434.38856	&-14.1572	&26.2 \\
55434.40002	&-14.2141	&26.1 \\
55434.41127	&-14.1490	&25.6 \\
55434.42196	&-14.1922	&25.3 \\
55434.43227	&-14.1831	&25.4 \\
55434.44326	&-14.2378	&25.8 \\
55434.45460	&-14.3113	&26.1 \\
55434.46798	&-14.4068	&27.1 \\
55434.47968	&-14.4386	&25.5 \\
55434.49150	&-14.4160	&25.5 \\
55434.50469	&-14.3966	&26.2 \\
55434.51904	&-14.4791	&27.4 \\
55434.53355	&-14.4316	&26.9 \\
55434.54732	&-14.5136	&25.9 \\
55434.55997	&-14.5404	&27.2 \\
55434.57210 &	-14.4825	&24.5 \\
55434.58393 &	-14.5221	&29.0 \\
55434.59768	&-14.5325	&35.7 \\
\hline
\end{tabular}
\end{table}
}

\section{Additional photometry}
\label{phot}

The observation of the spectroscopic transit of \8 b motivated the scheduling of photometric observations  of the transit, since the egress seemingly occurred several minutes before expectations on the 19 July 2010 sequence according to the ephemeris by \citet{l08}. Photometric observations to refine the ephemeris were therefore carried out from several sites in France and Spain on 25, 28 and 31 August, from both amateur and professional observatories, and are briefly described in Table \ref{pho}. Each lightcurve has been acquired and derived from observations independently; aperture photometry has been performed with optimized parameters for each system and conditions. Additionally, a transit light curve obtained on the night of September 13th by Ken Hose, was recovered from the Extrasolar Transit Database (ETD).

The conversion from UTC timestamps to Barycentric Dynamical Time (TDB) standard was done using the code by \citet{eastman2010}. Although not mentioned explicitly in the discovery paper, the KeplerCam timestamps were assumed to be provided in UTC and were therefore corrected to TDB by adding 65.184 seconds to the measured central times \citep[see][]{eastman2010}. Anyway, this correction does not change the ephemeris above the 1-$\sigma$ level.

Each light curve was fitted  individually  with the orbital and planetary parameters fixed to those by \citet{l08} in order to obtain the central times of transit. The error bars were estimated by means of the prayer bead method (see, for example, \citet{desert2009}), which consists in fitting a set of synthetic datasets obtained by shifting sequentially the residuals of the best-fit model rigidly and adding them back to the best-fit trapezoid function. The dispersion in the set of fit parameters obtained in this way is then used as an estimate of the uncertainties. This method permits to preserve the structure of the residuals and is therefore supposed to produce error estimates that take into account the covariant noise in the light curves. At each iteration of the processes, the limb-darkening coefficient corresponding  to each photometric band in which a transit was observed was drawn randomly from a normal distribution centred at the value obtained by interpolating the tables of \citet{claret2000} or \citet{claret2004} to the stellar parameters from \citet{l08}. The dispersion of this distribution is defined by the extremes values of the limb-darkening coefficients obtained by allowing the stellar parameters to vary within the reported 1-$\sigma$ intervals.
The KeplerCam lightcurves from the discovery paper \citep{l08} were fitted in a similar way. Figure \ref{phot} shows the light curves obtained by the different observers, together with the light curves from KeplerCam, binned to 10 minutes.

An improved ephemeris was obtained by linear regression of the complete set of central times obtained,  excluding the KeplerCam transit of November 1st 2007, since only the ingress was observed:
\begin{eqnarray}
\centering
&To [BJD-2\,450\,000]= 4437.67505 \pm 0.00042&\\
&P = 3.0763373 \pm 0.0000031\textrm{ days}&\;\;,
\end{eqnarray}
with a covariance $\mathrm{cov}(P,To) = -5.1443\times10^{-10}$, and reduced $\chi^2 = 0.75$. The value of $\chi^2$ probably means that the error bars of the central times of transits have been slightly overestimated. However, we decided not to scale them and therefore to report conservative ephemeris uncertainties.

This ephemeris is in good agreement with those obtained by \citet{simpson2010},  but the orbital period differs significantly (by almost 8-$\sigma$) from the value reported by \citet{l08}. As a consequence, the predictions for the July 19th transit differ by more than 19 minutes (8-$\sigma)$. Using our computation, the expected central time of transit observed by SOPHIE is
\begin{displaymath}
T_c[BJD-2\,450\,000]  = 5397.49230 \pm 0.00089\;\;,
\end{displaymath}
where the covariance has been considered for propagating the error from the ephemeris.

 To explore the discrepancy with the prediction by \citet{l08}, we adjusted the data of this target obtained by the HATnet telescopes together with the KeplerCam transits, in an attempt to reproduce the ephemeris reported in the discovery paper. The orbital period thus obtained is in agreement with our determination to the 2-$\sigma$ level and it differs from the period by \citet{l08} by more than 7 $\sigma$. We conclude there exists a problem --probably a misprint-- with the value reported in Table 13 of \citet{l08}, and that the observed discrepancy is not real\footnote{Curiously, the orbital period reported on Table 13 of the version of the paper found on  astro-ph (http://arxiv.org/pdf/0812.1161) differs from the published value. The astro-ph version of the orbital period is closer to our determination (they differ by 3.4 $\sigma$) and the difference in the corresponding predictions of the July 19th 2010 is about 6.6 minutes (less than 3 $\sigma$). This reinforces our conclusion that the discrepancy is not significant, but stems rather from a typographical error.}.

Using the above values for $To$ and $P$ we adjusted the phase-folded curve using the analytical formulae of \citet{mandelagol2002} with a linear limb-darkening law to obtain the radius ratio $k=R_p/R_*$, the system scale $a/R_*$ and the impact parameter $b = a/R_* \cos(i)$. The minimum value of the $\chi^2$ statistic was obtained using the down-hill simplex method \citep{neldermead65}, and the errors were estimated using the Prayer Bead method. At each iteration of the processes, the limb-darkening coefficient corresponding  to each photometric band in which a transit was observed was drawn randomly from a normal distribution centred at the value obtained by interpolating the tables of \citet{claret2000} or \citet{claret2004} to the stellar parameters from \citet{l08}. The dispersion of this distribution is defined by the extreme values of the limb-darkening coefficients obtained by allowing the stellar parameters to vary within the reported 1-$\sigma$ intervals.

The obtained parameters are  in perfect agreement with those from \citet{l08}, although with a larger uncertainty. We thus use the transit parameters of the latter paper to model  the spectroscopic transit.

\begin{figure}
\epsfig{file=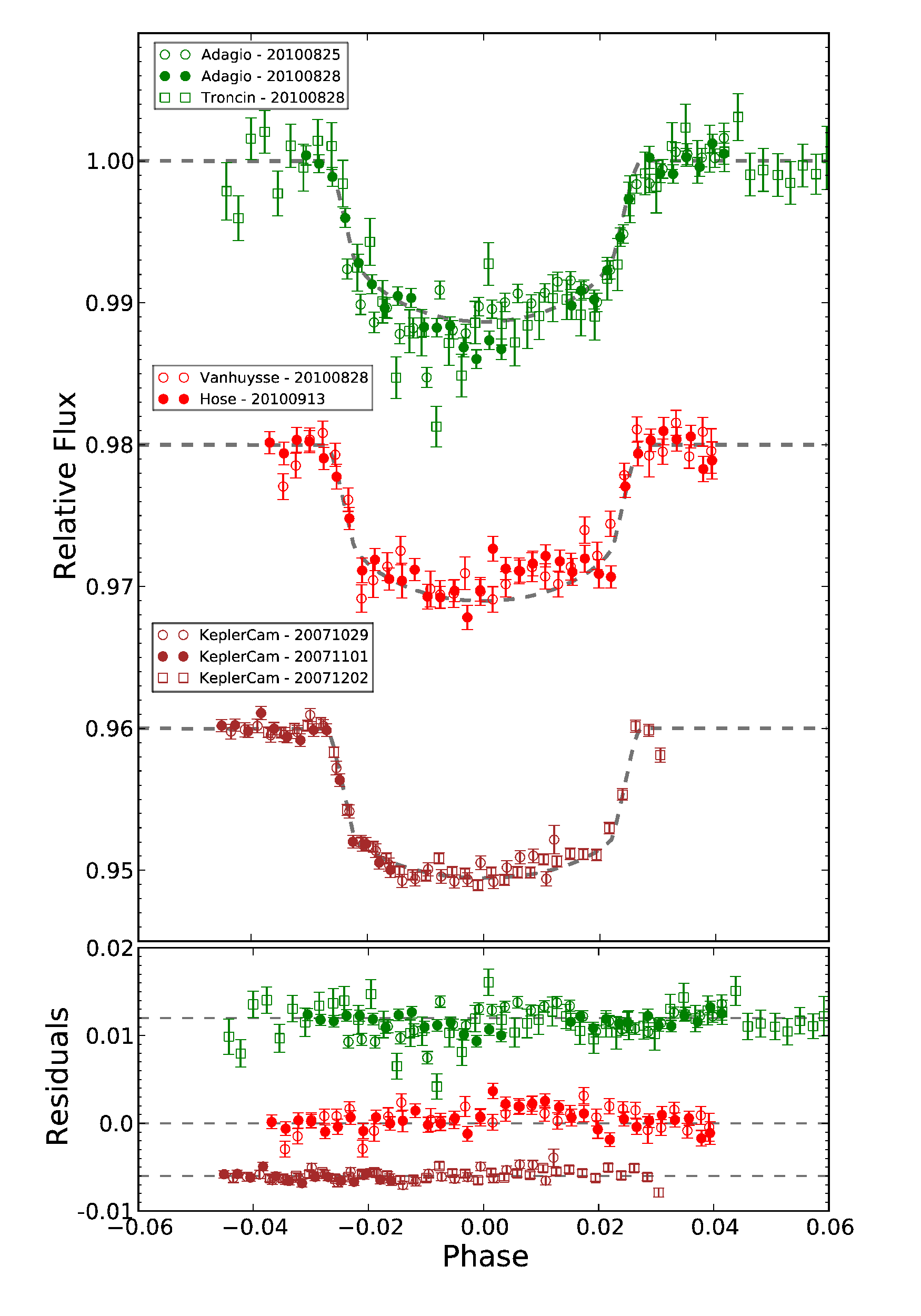,width=0.5\textwidth}
\caption{Upper panel: Transit light curves of HAT-P-8 in three different bands: V (top; green), R (middle; red) and Sloan-\emph{z} (bottom; brown), shifted vertically to improve legibility. Different symbols indicate different observers or nights, as indicated in the legend. The dashed curve is the best fit model for each photometric band. Bottom panel: Residuals to the best fit model. The data has been shifted to avoid crowding.}
\label{phot}
\end{figure}

\section{Modeling the spectroscopic transits}
\label{mod} 

\begin{figure}
\begin{center}
\vspace*{-3cm}
\epsfig{file=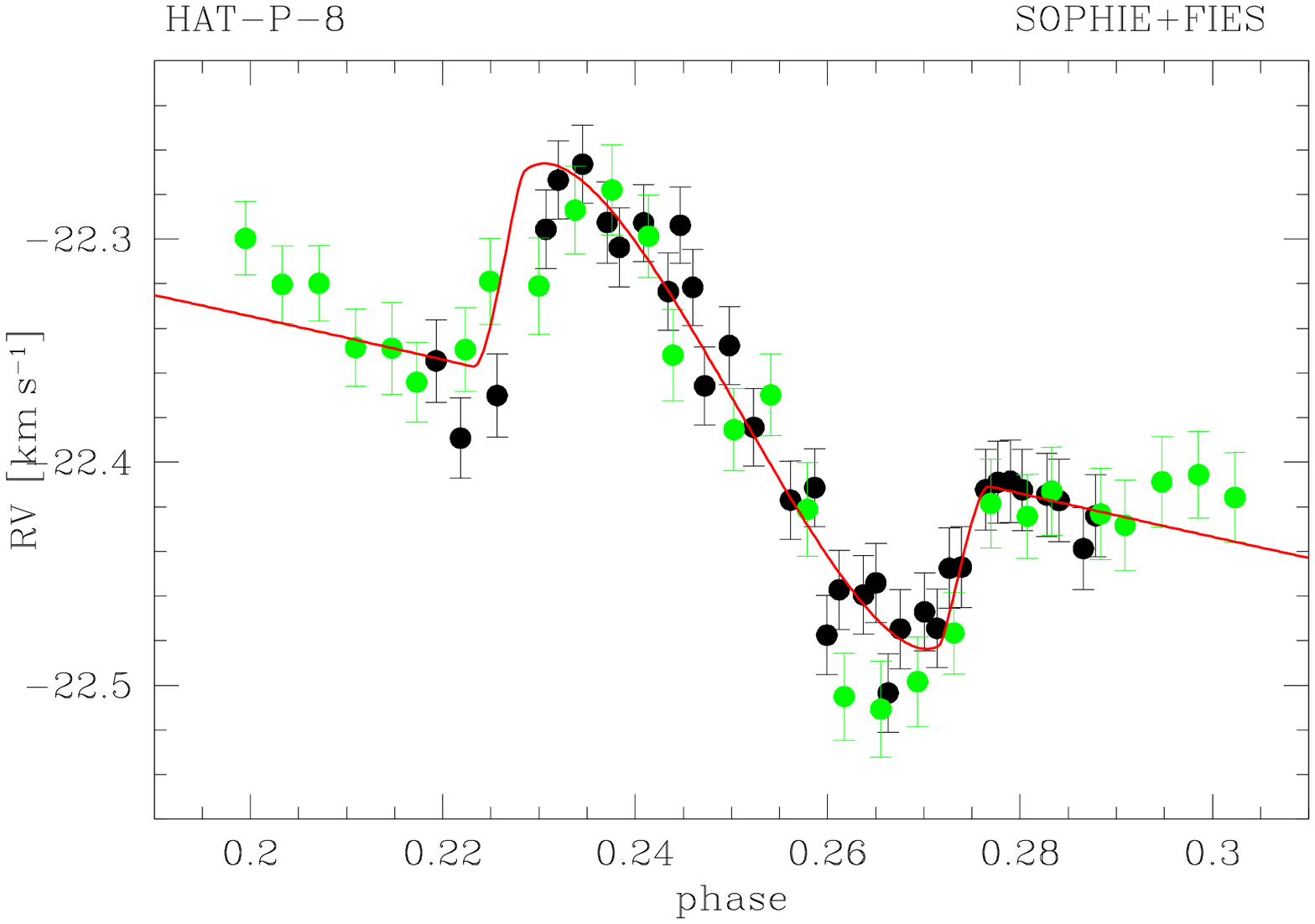,width=0.4\textwidth,angle=0}
\vspace*{-2cm}
\vspace*{-3cm}
\epsfig{file=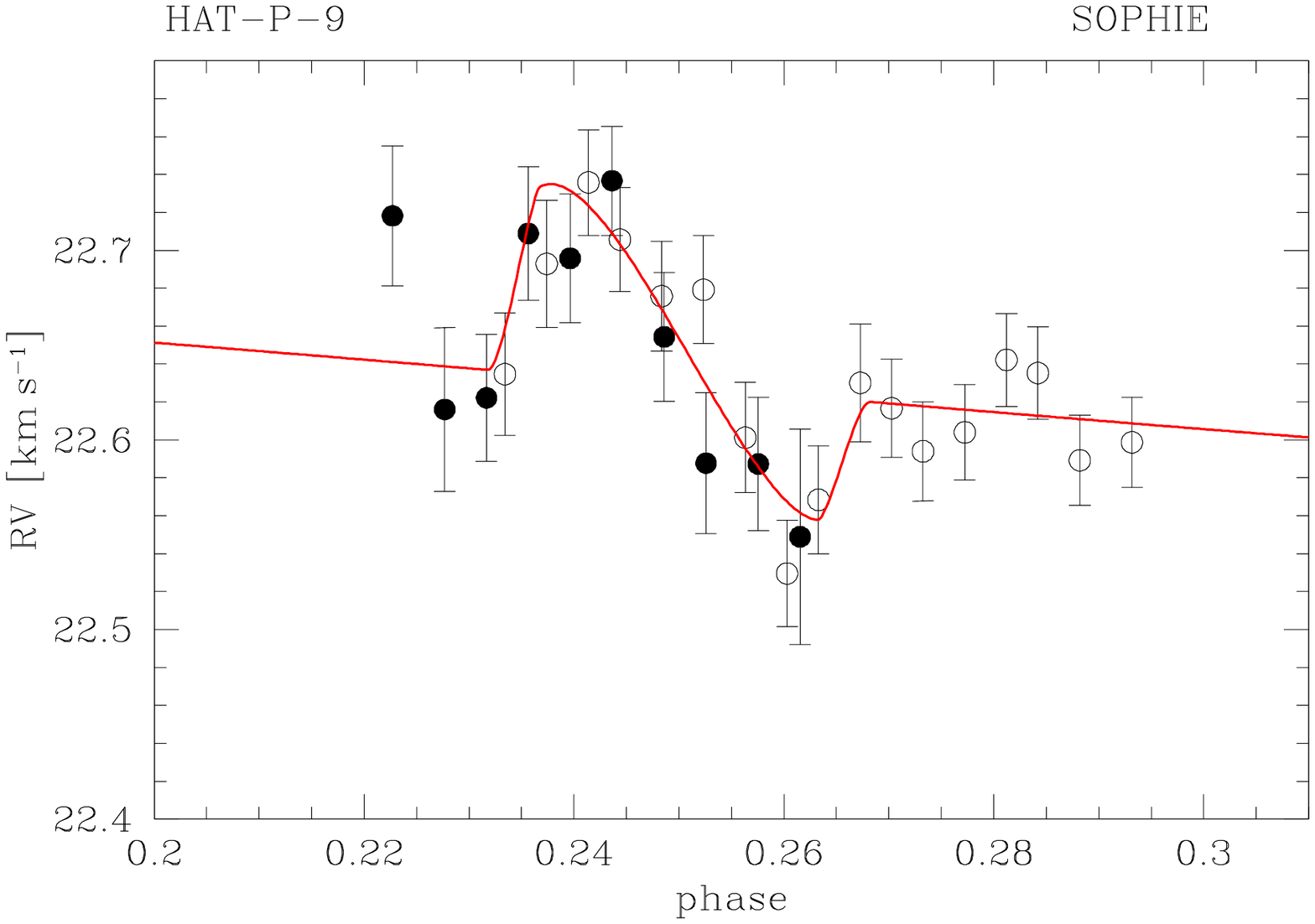,width=0.4\textwidth,angle=0}
\vspace*{-2cm}
\vspace*{-3cm}
\epsfig{file=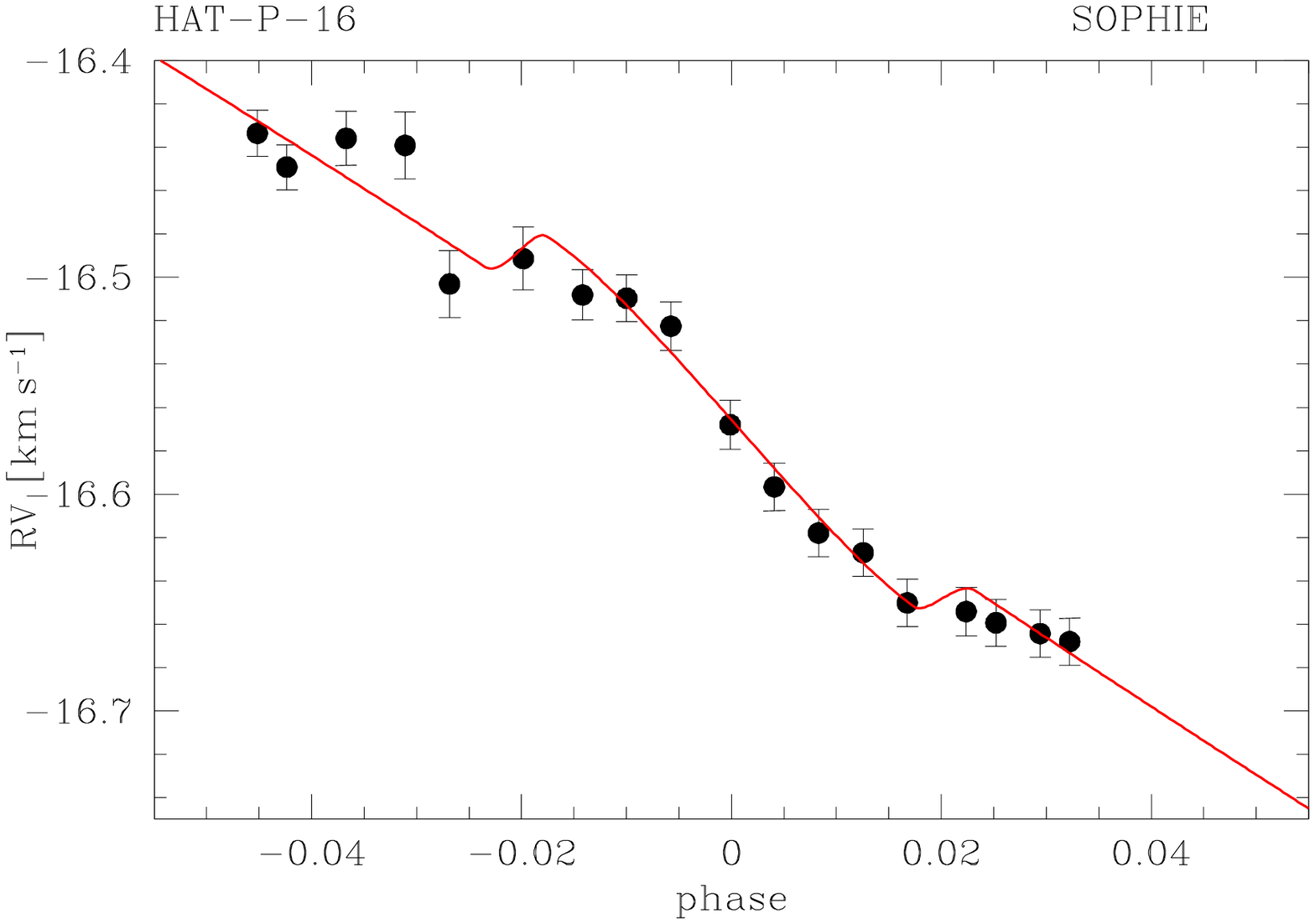,width=0.4\textwidth,angle=0}
\vspace*{-2cm}
\vspace*{-3cm}
\epsfig{file=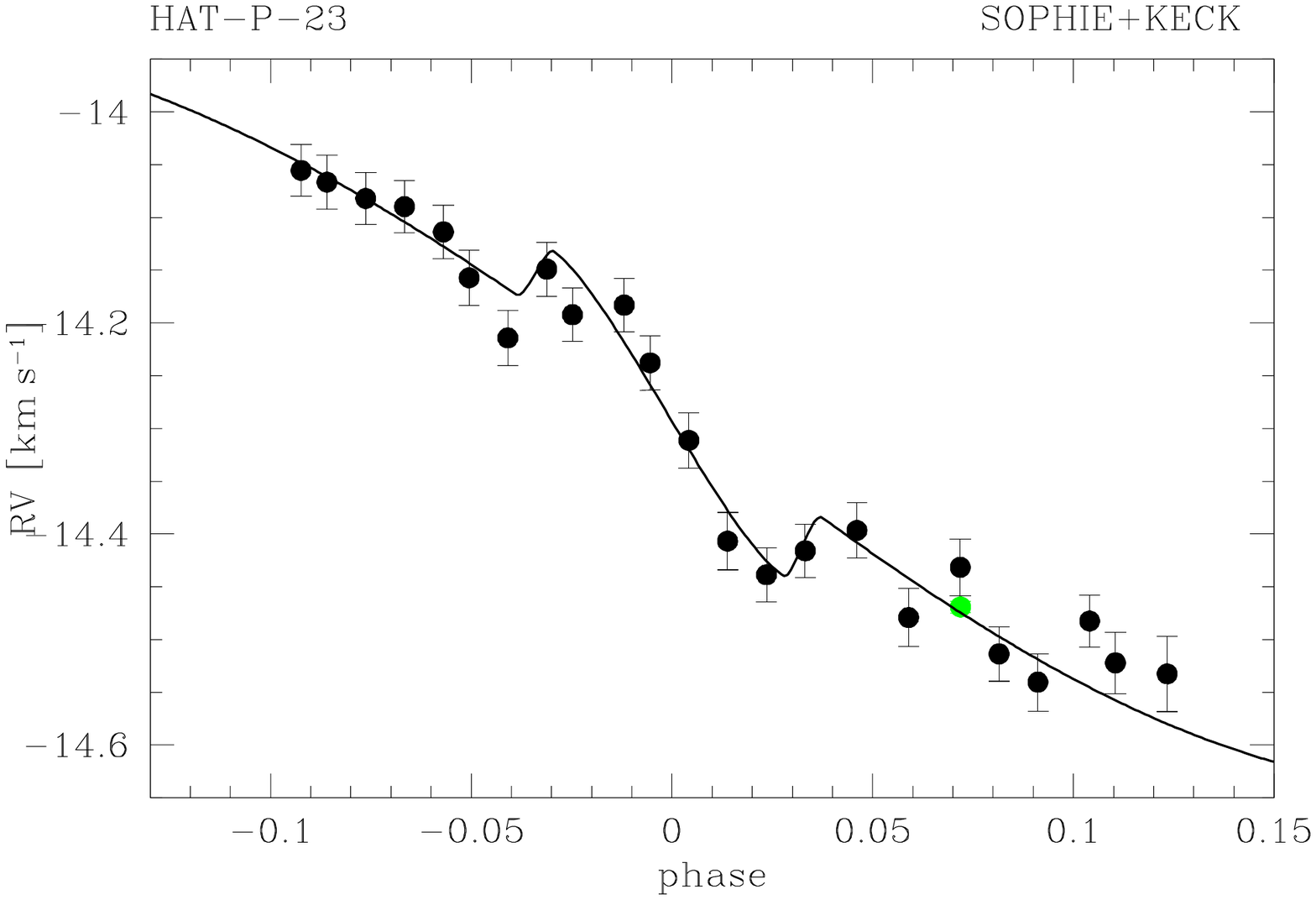,width=0.4\textwidth,angle=0}
\vspace*{-2cm}
\caption{Close-in view of the radial-velocity measurements of \8, \9, \6 and \2 obtained with SOPHIE during transit nights. The best-fit Rossiter-McLaughlin models are superimposed as a red dashed line. FIES data of \8 are included as green circles \citep{simpson2010}. Two separate observing nights are shown with different symbols for \9 (filled circles: 28 December 2010, open circles: 1 January 2011).}
\label{rm}
\end{center}
\end{figure}

Modeling of the radial-velocity anomaly has been performed using the analytical approach developed by \citet{ohta05}.  The complete model has 10 parameters: 
the four standard orbital parameters for circular orbits, the radius ratio $R_p/R_*$,  the orbital semi-major axis  to stellar radius $a/R_*$ (constrained by the transit duration), the sky-projected angle between the stellar spin axis and the planetary orbital axis $\lambda$, the sky-projected stellar rotational velocity \vsini, the orbital inclination $i$, and linear 
limb-darkening coefficients based on \citet{claret2004} tables for 
filter $g$'. Our goal here is to measure the spin-orbit projected angle $\lambda$
from the fits of the RM anomaly shape; we thus let free this parameter  $\lambda$ 
in the fits. As in the case of HAT-P-6 \citep{hebrard11}, the parameters having 
a significant impact on the RM fits and the $\lambda$ value are mainly the projected stellar 
velocity $v \sin i_s$ and the systemic velocity $\gamma$ of the data,  and less critically, the RV semi-amplitude $K$ and orbital inclination $i$. The $K$ and $i$ parameters were fixed to their mean value in the fit but we explored the range of their error bars and quadratically added their small contribution to the final errors on \vsini and $\lambda$.
 We constructed a grid scanning the different values for $\lambda$, $v \sin i_s$, and the systemic velocity $\gamma$, and computed the $\chi^2$ for each model. We performed a conservative re-scaling of the  $\chi^2$ array, introducing a correction for the $\chi^2$ fluctuations due to the limited number of degrees of freedom, as explained in \citet{hebrard02}. Adopting the minimum $\chi^2$ obtained for each value on the grid of a given parameter (e.g. $\lambda$), we determined the uncertainty on this parameter from the $\chi^2$, $\Delta \chi^2$ variations of 1, 4, and 9 determining the interval confidences at 1, 2, and 3 $\sigma$. 

 In RM fits, the degeneracy between $v \sin i_s$ and $\lambda$ increases with the orbital inclination $i$. For impact parameters $b$ that are not too small this correlation is taken into account in this work, by exploring the
entire space of possible values for $v \sin i_s$ and $\lambda$.
Figure \ref{rm} shows a zoomed view of the spectroscopic transits and their best-fit solutions.  The confidence interval contours estimated from rescaled-$\chi^2$ variations against $\lambda$ and $v \sin i_s$ are plotted in Fig. \ref{chi}.

\subsection{\8}
We used the orbital parameters as derived in Section 2.1 and 3 and the transit parameters from \cite{l08}. 
Two datasets were used in search for the best solution: first, the SOPHIE data set (35 measurements during the transit night), as described in Section 2.1; second, the data set composed of the SOPHIE data and the FIES measurements as recently published by \citet{simpson2010}. This latter study includes a total of 28 data points. When quadratically adding 15 m/s to the FIES individual errors (as in \citet{simpson2010}), and allowing for a free offset between both SOPHIE and FIES data sets, we obtain values of \vsini = 14.5  $\pm$  0.8 km/s,  $\lambda$ = -17$^\circ$  $\pm$  8 $^{\circ}$ for the projected rotational velocity and spin-orbit angle, respectively, with a reduced $\chi^2$ of 1.19. 
 Fits with the extreme values of $i$ show that an additional 
 uncertainty should be quadratically added to the uncertainty obtained from the
$\chi^2$ variations. 
We checked that, as expected, variations of other parameters within
their error bars have no significant impact on $\lambda$. The final value of  $\lambda$ is -17$^\circ$$^{+9.2}_{-11.5}$.

The values reported by \citet{simpson2010} are -9.7$^{\circ}$ for  $\lambda$ (1-$\sigma$ range from -17.4$^{\circ}$ to -0.7$^{\circ}$) and 11.8 $\pm$ 0.5 km/s for \vsini. Adding the SOPHIE values therefore tends to slightly increase the spin-orbit angle (both estimates being in agreement within 1-$\sigma$), while the derived \vsini\  increases by about 2-$\sigma$. It is a known consequence in fitting spectroscopic transits of stars that are significantly rotationally broadened, that the \vsini\ is over-estimated \citep{hirano10}, with consequences on the estimated $\lambda$ value that are of order of magnitude smaller than our error bar \citep{triaud09}. Figure \ref{chi} shows the distribution of $\chi^2$ values against $\lambda$ for this system.

\subsection{\9}

We first fit the planetary orbit, using the full set of SOPHIE data since 2007, and new ephemeris by \citet{dittmann10}: we find less dispersion in the residuals and a planetary mass slightly smaller than given in \citet{shporer09}. The RV best-fit semi-amplitude is K=73.8$\pm$ 6.7 m/s and \mp = 0.67 $\pm$ 0.08 \Mjup   with  M$_s$ = 1.28$\pm$0.13 \Msun.

For this planet, the semi-amplitude of the RM anomaly is larger than the amplitude of the orbital motion, a rare situation again explained by the large planetary radius, large stellar rotational velocity, and relatively light planet (hence small orbital amplitude).
The Rossiter effect is then modeled by combining both December 2010 and January 2011 sequences, and letting a free offset between these series. Half of the transit is observed twice, so that the offset is determined from a subset of about eight measurements per series. The offset is found to be 37 m/s in order to minimize the residuals. With the total sample of 27 measurements during or close to the transit, the best-fit of the Rossiter-McLaughlin anomaly gives the following solution: rotational velocity of 12.5  $\pm$ 1.8 km/s (in agreement with \citet{shporer09}) and spin-orbit angle $\lambda$ of -16$^\circ$  $\pm$ 8$^\circ$. The reduced $\chi^2$ value is 0.68.  Taking into account the impact of RV semi-amplitude and inclination errors does not change significantly the value nor its error bar. The transit data and its best fit are shown in Figure \ref{rm}, and the $\chi^2$ distribution of the derived quantity is shown on Figure \ref{chi}.

\subsection{\6}
\begin{figure}
\begin{center}
\vspace*{-2.5cm}
\vspace*{-3cm}
\epsfig{file=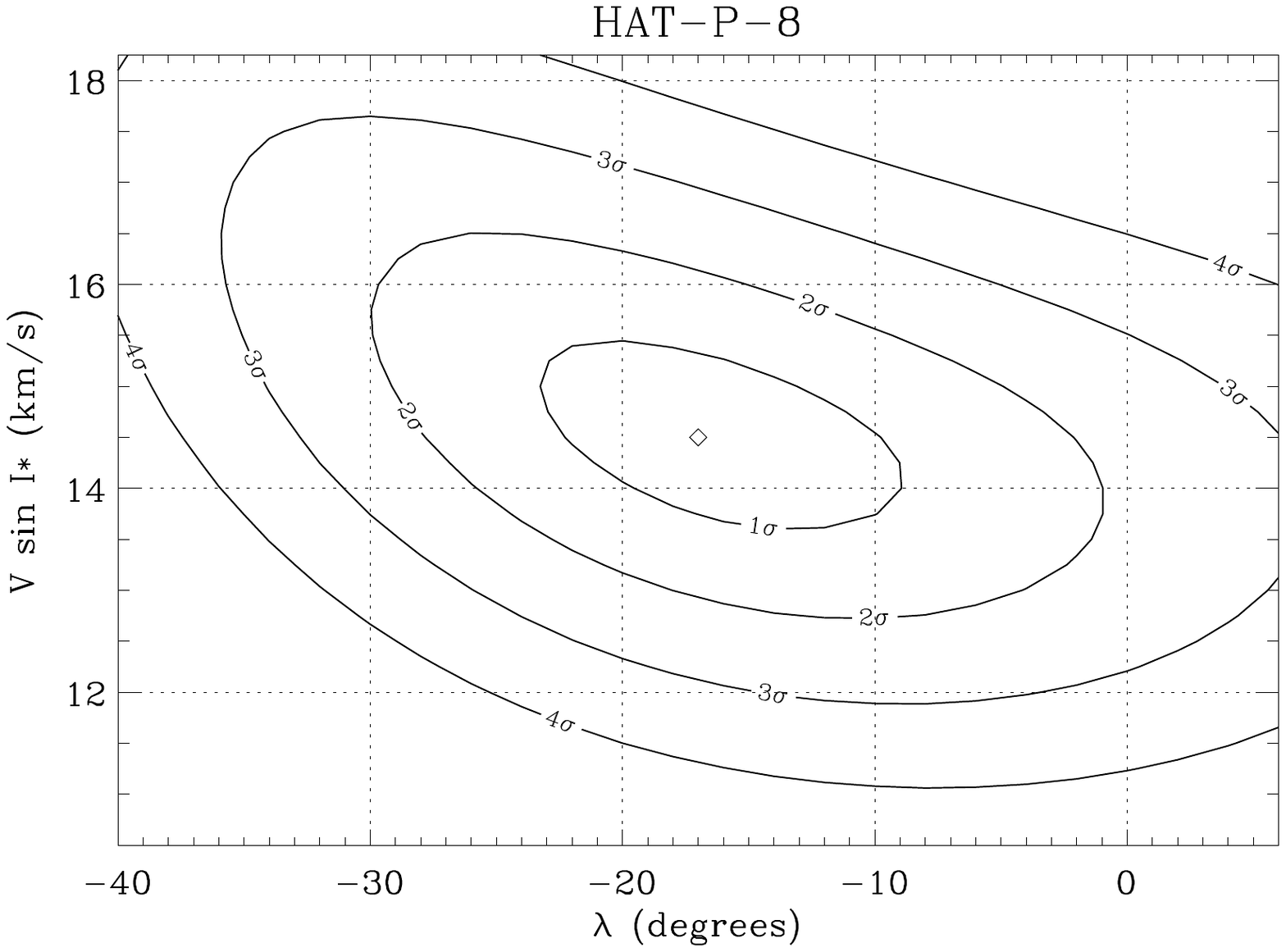,width=0.45\textwidth}
\vspace*{-2.5cm}
\vspace*{-3cm}
\epsfig{file=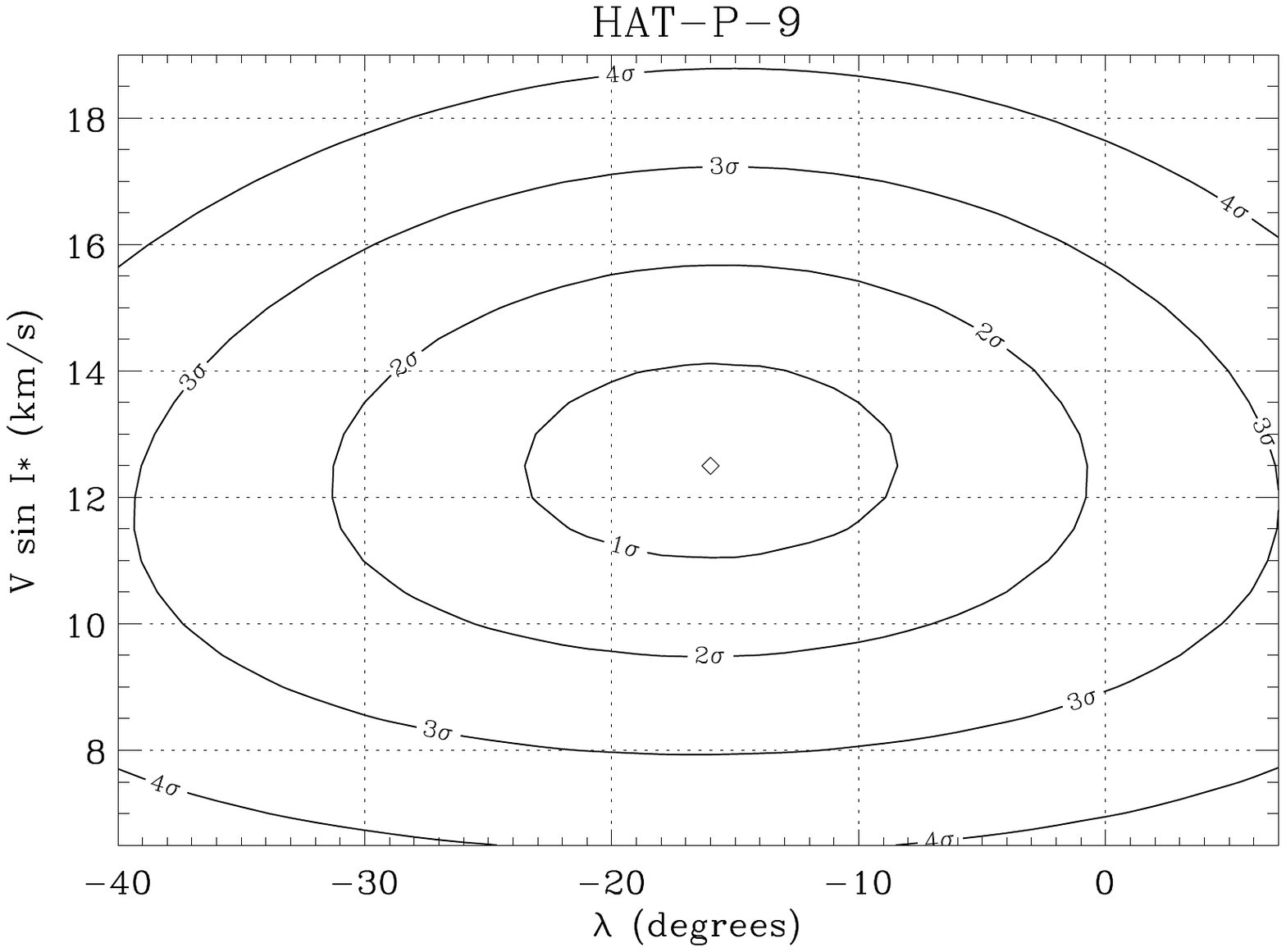,width=0.45\textwidth,angle=0}
\vspace*{-2.5cm}
\vspace*{-3cm}
\epsfig{file=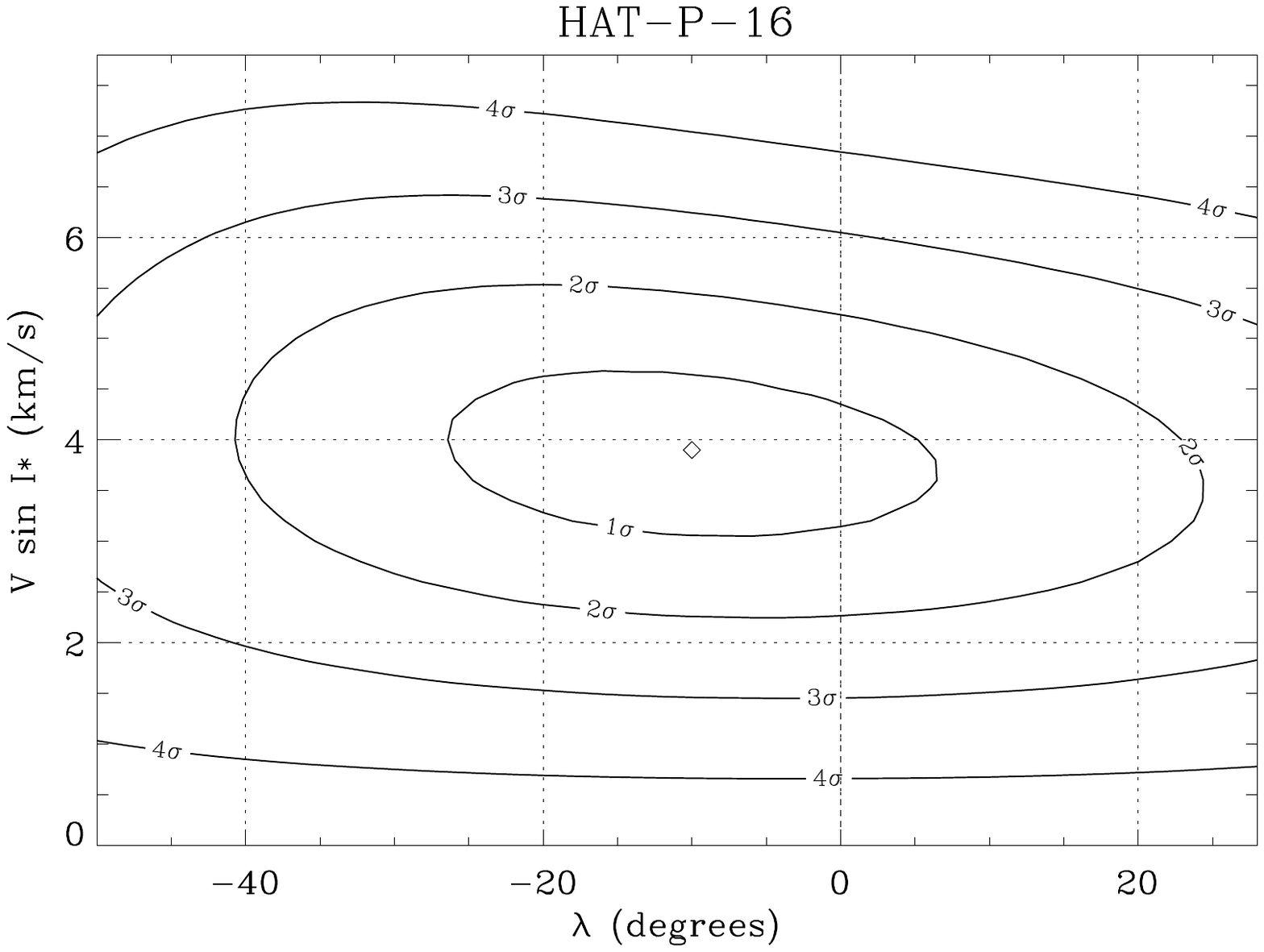,width=0.45\textwidth,angle=0}
\vspace*{-2.5cm}
\vspace*{-3cm}
\epsfig{file=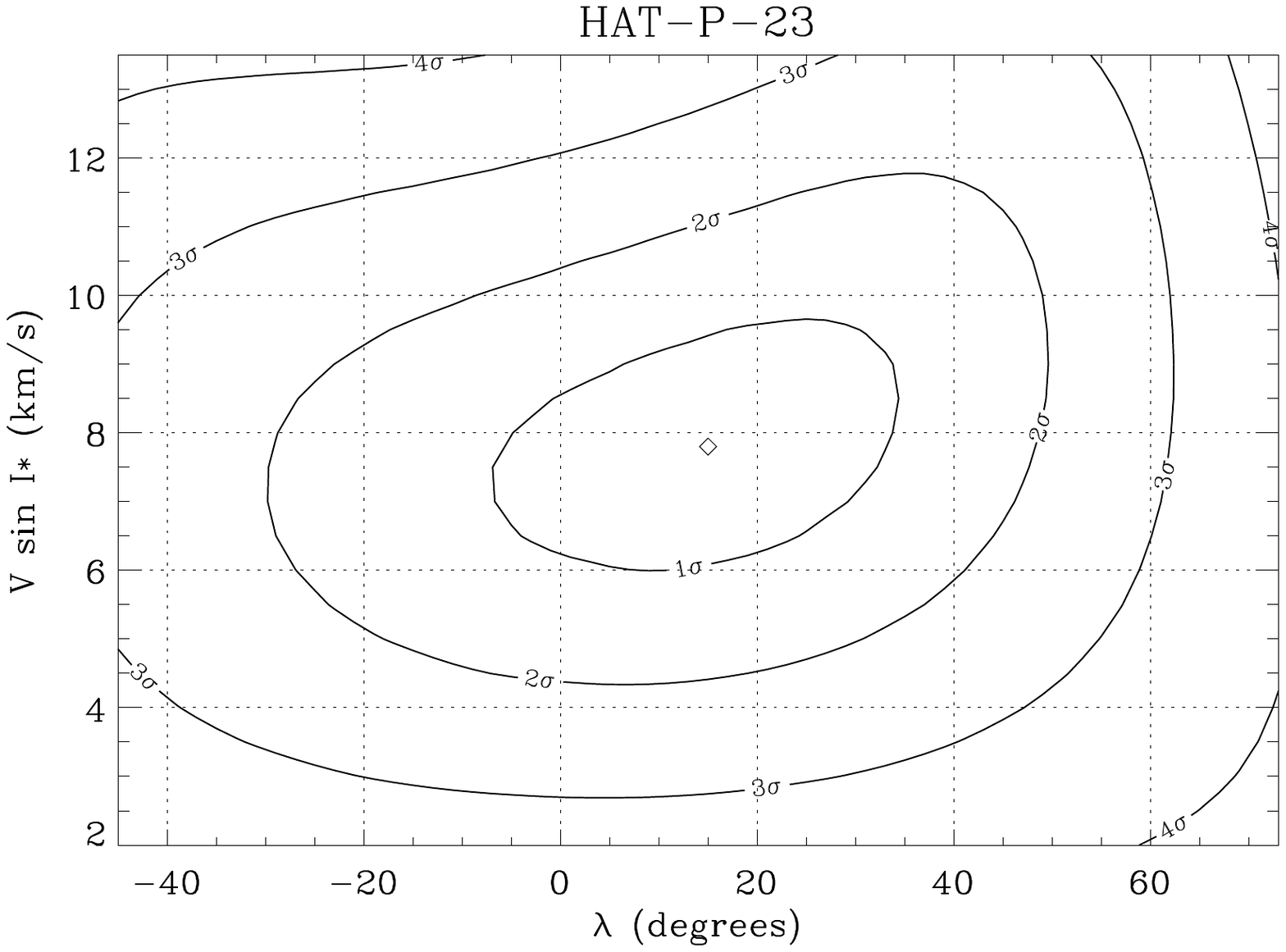,width=0.45\textwidth,angle=0}
\vspace*{-0cm}
\caption{$\chi^2$ contour plot for the $\lambda$ values from the Rossiter-McLaughlin fits.} 
\label{chi}
\end{center}
\end{figure}

Due to a lower value of the stellar rotational velocity and a non-inflated planet, the amplitude of the Rossiter anomaly on the star \6 is small, while the individual measurement precision is greater compared to the other sequences.
We perform the fit using the transit parameters as derived by \citet{buchhave10} and find a best-fit solution for a stellar projected rotational velocity of 3.9 $\pm$ 0.8 km/s (in agreement with \citet{buchhave10}) and a spin-orbit angle of -10 $\pm$ 16 $^\circ$. The reduced $\chi^2$ of the fit is 0.89, and the fitted data and $\chi^2$ distribution are shown on Figures \ref{rm} and \ref{chi}.  We checked that the variation of the RV semi-amplitude and inclination within their error bars did not have a noticeable impact on the value and error of the spin-orbit angle and rotational velocity in this case. Note that with an effective temperature of 6158 K, one may expect the stellar rotational velocity to be greater than 3.9 km/s. An average value for such a star would rather be about 10 km/s \citep{nordstrom04}. It is thus possible that the stellar rotation axis of \6 be inclined over the line of sight, resulting in a real spin-orbit angle larger than measured.

\subsection{\2}
Finally, the SOPHIE data of \2 were adjusted in the same way than previously, using the transit parameters and ephemeris of \citet{bakos2011b}.
We derive a stellar rotational velocity of 7.8 $\pm$ 1.6 km/s and a projected angle between the orbital plane and the stellar equatorial plane, of 15$^\circ$ $\pm$22$^\circ$.  Again, we checked how the variations of the RV semi-amplitude and inclination within their error bars would affect the value and error of the spin-orbit angle and rotational velocity; here the error on the inclination has some impact, which is included in the final error bars.
The \vsini\ derived from the fit is in good agreement with the value reported in \citet{bakos2011b} from stellar analyses.
Figure \ref{chi} shows the $\chi^2$ distribution as a function of $\lambda$.

\section{Discussion}
\label{disc}

Table \ref{result} summarizes the values derived by our analyses of the spectroscopic transits of \8 b, \9 b, \6 b, and \2 b.
The total number of systems with published projected spin-orbit angles is now 37. Ten of them are in the range $\ | \lambda \ | =$ 20-120$^\circ$, 7 are retrograde (more than $\sim$120$^\circ$), 22 are less than $\sim$20$^\circ$, among which the four systems presented here. Let us recall that the $\lambda$ value is the projected spin-orbit angle. The analysis of  the real spin-orbit angle $\psi$ distribution by \citet{triaud10} showed that 45 to 85 \% of hot Jupiters may in fact have an orbital inclination greater than 30$^\circ$. The recent study by \citet{schlaufman} discusses the projected stellar rotation axis and the spin-orbit measurements and  concludes that systems of massive stars and planets are more likely to show an apparent misalignment.

Since the discovery of the first misaligned system \citep{hebrard08}, there have been a lot of processes invoked for explaining them: three-body Kozai mechanism \citep{fabrycky07}, planet-planet scattering \citep{chatterjee08, ford08}, tilted disks \citep{bate10,lai10}, and insufficient tidal torque of stars hotter than 6200 K -hence a conservation of an original misalignement due to the preceeding mechanims for planets around hot stars \citep{winn10}.  
Yet another process should be taken into account for this global picture: the tidal instability, which may affect all systems by tilting the rotation axis of the star in relatively short timescales \citep{cebron10}, and thus contribute to the distribution of orbital inclinations. The four systems analysed in this paper seem to be out of the so-called forbidden zone -- which is defined by an orbital period less than 1/3 of the stellar rotational period \citep{lebars10}, and where the tidal instability cannot grow --  and thus might have suffered a change of obliquity during their history. Complete simulations of the tidal evolution of the systems should be carried out in order to unveil possible impact of the tides onto the stars or the planets, particularly \6 and \2 which exhibit non-zero eccentricity despite their age (2.0 and 4.0 Gyr respectively).

It is thus almost easier now to explain misaligned systems, than the more abundant hot Jupiters  close to alignment. The addition of four systems  not strongly misaligned, as well as other recent discoveries renews our interest for this majority of systems that are "surviving" the vast panel of possible causes for misalignment. Why should hot Jupiters get aligned with the equator of their star, and which parameter(s) would permit to predict the behaviour of the orbital planes? What constraint do we have on the sample of the aligned hot Jupiters, in terms of exclusion of other planets or stellar companions in the system, or stellar and disk properties?

\begin{figure}
\begin{center}
\epsfig{file=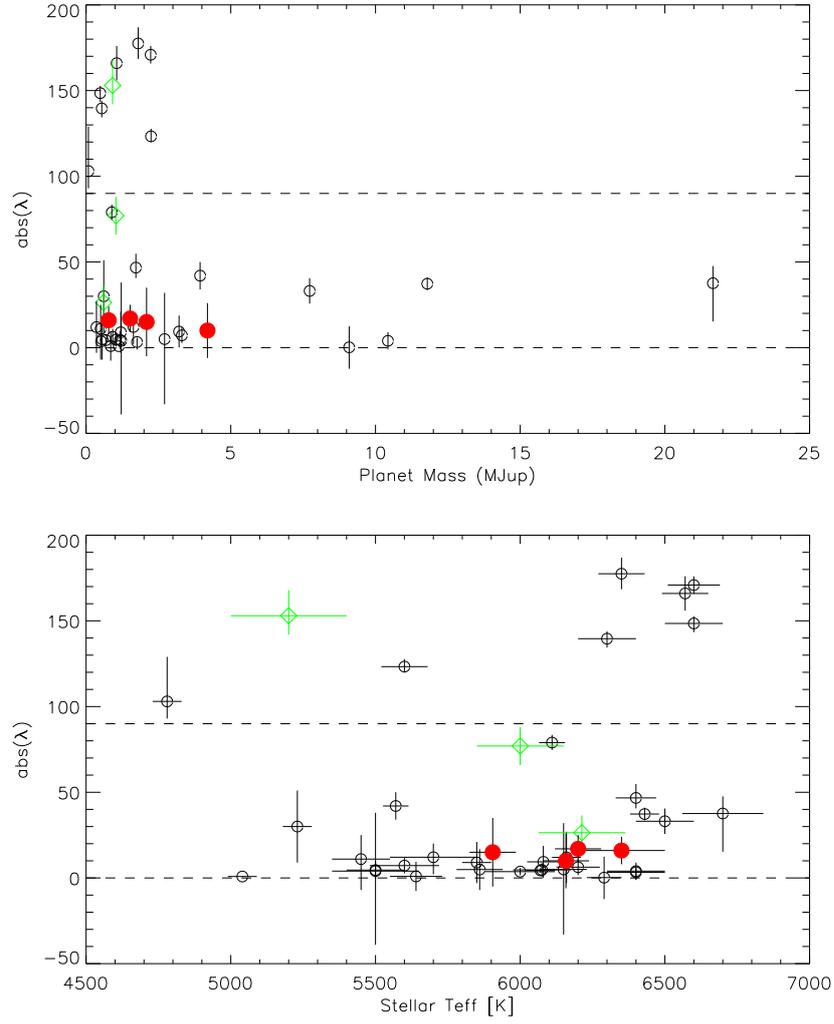,width=0.75\textwidth,angle=0}
\caption{Absolute value of the spin-orbit angle as a function of planetary mass (top) up to 25 \Mjup\ and stellar effective temperature (bottom). The red filled large symbols show the new measurements.   The uncertain values for CoRoT-1b, WASP-2 and Kepler-8 are included as green losenges.}\label{stat}
\end{center}
\end{figure}

Figure \ref{stat} (top) shows the distribution of the absolute value of the $\lambda$ projected spin-orbit angle with respect to the planetary mass, updated from a similar plot by \citet{hebrard10} with almost a factor of two more systems. 
It shows that most of the systems show a low projected spin-orbit angle over the mass range of Jupiter-like planets. Among the planets in this mass range with misaligned orbits are included planets in binary systems (as WASP-8, \citet{queloz10}) and seven systems in retrograde orbits ($\ | \lambda \ | >100^\circ$). These latter systems show some dependency with the planetary mass:  only planets less massive than 3 \Mjup\ are retrograde.

There are at the moment: i) notable exceptions to any general rule (e.g. HAT-P-11 recently announced by \citet{winn10b}, a $\lambda$=103$^\circ$ orbit around a cool star, depicting an effect linked to the tidal dissipation rather than the sole effective temperature of the star) and ii) partial knowledge of the systems (unknown other planets or stellar companion in the system), that limit our understanding of the common behaviour.  Figure \ref{stat} (bottom) shows an updated version of the $\ | \lambda \ |$ dependency with stellar effective temperature, as in \citet{winn10}. Our data might add a new exception of the rule that hot stars have planet with high obliquities (\9 is has a  low-angle misalignment, compatible at 2-$\sigma$ with zero with T$_{eff}$=6350K), although it may also indicate that the obliquity never changed during the system's evolution. Additionally, two of the systems reported in this paper lie close to the limiting temperature proposed by \citet{winn10b}, \8 and \6 (with T$_{eff}$=6200 and 6158 K respectively). Finally, \2 with a cooler star (T$_{eff}$=5905K) and aligned orbit stands within the majority of such systems. The current sample of planets with measured spin-orbit projection angle hinders statistically significant conclusions concerning the difference between planets orbiting hot and cold host stars from being drawn.
Indeed, considering T$_{eff}$= 6250 K as the limiting temperature between samples, a Kolmogorov-Smirnov test on the data fails at rejecting the null hypothesis (i.e. that the sample of measured $\lambda$ for hot and cold systems are drawn from the same distribution) with a significance below $\alpha = 0.11^{+0.17}_{-0.09}$, where the error bars are issued from 10000 bootstrap realisations of the data, using the uncertainties of effective temperature and $\lambda$ reported in the literature. This means that at the 1-$\sigma$ level, there exists a 28\% of chance of erroneously rejecting the hypothesis that both populations are one and the same. Interestingly, we found slightly lower values for the significance when the limiting Teff is slightly higher: for Teff between 6300 K and 6450 K, the null hypothesis can be rejected at the 1-$\sigma$ level with a significance $\alpha=0.2$. We note, however, that for these temperatures only about 10 systems are classified as orbiting hot stars.

The first steps are thus to systematically complete our understanding of the systems, both by following up known transiting hot Jupiters in radial velocity on the long term, and by deriving detection limits of stellar companions by direct imaging. Companions able to efficiently misalign the planetary systems through Kozai oscillations should generally be separated by no more than $\sim$100 AU from the stars hosting the planets, otherwise the Kozai period exceeds that of the relativistic precession, which then damps the Kozai cycles \citep{fabrycky07}. Unfortunately, the transiting systems discovered during surveys such as HATnet or SuperWASP are commonly located at large distances ($>$ 200 pc). Consequently, the adaptive optics imaging surveys, such as the ones carried out at Subaru \citep{narita10} or VLT (Ehrenreich et al., in prep.) have to probe the stellar environment within a few 0.1 arcsec. Searching for low-mass stars or brown dwarfs at these separations requires using innovative techniques such as angular differential imaging or sparse aperture masking (Lacour et al. 2011, in prep.). Alternatively, such companions could be detected through a long-term drift in the radial velocity of the star, providing they are observed long enough.
A systematic collection of such data would provide useful constraints to the multi-body mechanisms and thus reinforce all other proposed scenarios for single planets around single stars.  

\onecolumn
\begin{table}
\centering
\caption{Photometric observations of the transits of \8 b used to refine the ephemeris.}
\label{pho}
\begin{tabular}{lcccc}
\hline
Observer & Location & Instrument & Photometric Band & Date of transit obs.\\
\hline
Assoc.~Adagio    & Toulouse, F & 82-cm, SBIG STL-6303E & V& 08.28.10 and 08.31.10\\
Hose&Portland, Oregon& 32-cm, QSI 516wsg&R& 09.13.10\\
Troncin    & OHP, F & 120-cm Tektronix TK1024& V& 08.28.10 \\
Vanhuysse & La Palma, S & 35-cm, SBIG STL-1001e& R& 08.28.10 \\
\hline
\end{tabular}
\end{table}

\begin{figure}
\begin{center}
\epsfig{file=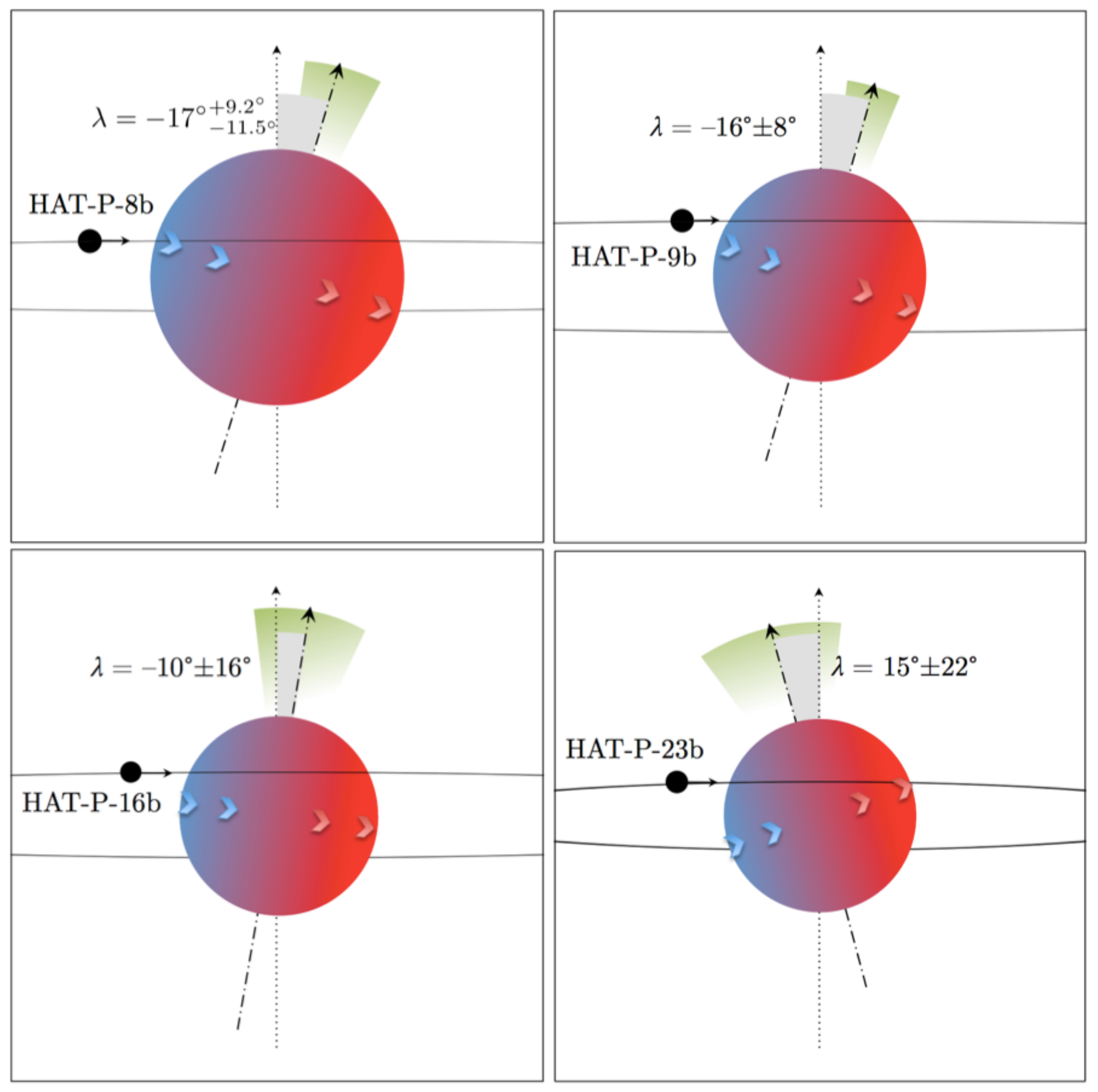,width=0.75\textwidth}
\caption{Sketch of the orbital and rotational orientations for all four systems discussed here, as viewed from Earth.}
\label{sketch}
\end{center}
\end{figure}

\begin{table}
\begin{center}
  \caption{Spectroscopic transit modeled parameters of \8, \9, \6, and \2.  Effective temperatures and planetary masses are taken from the discovery papers, except the planet mass of \9\, which is re-estimated in this work.}
  \label{result}
\begin{tabular}{lcccccc}
\hline
\hline
Star 			& V0		&\vsini	&	$\lambda$ & $\chi^2_{red}$ & T$_{eff}$ & \mp\\
			& km/s	& km/s	& degrees & &K	&\Mjup \\
\hline
\8			&-22.384 $\pm$ 0.0035		&	14.5 $\pm$0.8		&	-17 (-11.5,+9.2)	& 1.19& 6200$\pm$80&	1.52$\pm$0.18\\
\9			&+22.628 $\pm$ 0.008		&	12.5 $\pm$1.8	&	-16 $\pm$ 8	& 0.68&6350$\pm$150&0.67$\pm$0.08\\
\6			&-16.553 $\pm$ 0.0035		&	3.9 $\pm$ 0.8		&	-10.0 $\pm$ 16	& 0.89&6158$\pm$80&4.193$\pm$0.094\\
\2			&-14.259 $\pm$ 0.007		&	7.8 $\pm$ 1.6		&	+15 $\pm$ 22 & 1.02&5905$\pm$80&2.09$\pm$0.11\\
\hline
\end{tabular}
\end{center}
\end{table}
\twocolumn

\acknowledgements{Thanks to the referee Josh Winn for his very helpful remarks that improved the paper a lot. We gratefully acknowledge the Programme National de Plan\'etologie (telescope time attribution and financial support), the Swiss National Foundation and the Agence Nationale de la Recherche (grant ANR-08-JCJC-0102-01) for their support. D.E. is supported by CNES. A warm thank to the OHP staff, for their great care to optimise the observations. NCS acknowledges the support by the European Research Council/European Community under the FP7 through Starting Grant agreement number 239953. NCS also acknowledges the support from Funda\c{c}\~ao para a Ci\^encia e a Tecnologia (FCT) through program Ci\^encia\,2007 funded by FCT/MCTES (Portugal) and POPH/FSE (EC), and in
the form of grants reference PTDC/CTE-AST/66643/2006 and PTDC/CTE-AST/098528/2008. AE is supported by a fellowship for advanced researchers from the Swiss National Science Foundation. 
}

\bibliographystyle{aa}
\bibliography{references}

\end{document}